\documentclass[a4paper,11pt]{article}
\usepackage{snellen,graphicx,cite}
\usepackage{times}
\begin{document}
\baselineskip=13pt
\title{YOUNG EXTRAGALACTIC RADIO SOURCES}
\author{Ignas Snellen}
\address{Institute of Astronomy\\
Madingley Road\\
Cambridge, CB3 0HA\\
United Kingdom\\
E-mail:snellen@ast.cam.ac.uk}
\author{Richard Schilizzi}
\address{Joint Institute for VLBI in Europe\\
Postbus 2\\
7990 AA, Dwingeloo\\
The Netherlands\\
E-mail:rts@jive.nfra.nl}
\maketitle
\abstract{Gigahertz Peaked Spectrum (GPS) sources and Compact 
Symmetric Objects (CSO) are selected in very different ways, but have
a significant overlap in properties. Ever since 
their discovery it has been speculated that they are young 
objects, but only recently, strong evidence has been provided indicating that 
GPS sources and CSOs are indeed the young counterparts of large, 
extended sources. They are therefore the objects of choice to 
study the initial evolution of extragalactic radio sources.
Observational constraints on the luminosity evolution of young radio 
sources have mainly come from number density statistics and
source size distributions, 
indicating that young sources should decrease in luminosity by a factor
$\sim 10$ as they evolve to extended objects. 
We argue that the growth or decay in radio power of the individual 
objects has a strong influence on the slope of their collective luminosity 
function. This has
led to a new method of constraining the evolution of young sources
by comparing their luminosity function to that of large extended 
objects. The luminosity function of GPS sources is shown to be flatter than 
that of extended objects. This is consistent with the proposed scenario
that young sources increase in radio luminosity and
large size radio sources decrease in luminosity with time, but larger
homogeneous samples are needed.
The new large area radio surveys and 
the current sensitivity and flexibility of VLBI networks allow the 
construction and investigation of such large and homogeneous samples of young 
sources.}

\section{Selection of Young Radio Sources: GPS versus CSO}

The identification and investigation of the young counterparts of 
`old' extended radio sources is a key element in the study
of the evolution of radio-loud active galactic nuclei.
Two classes of compact radio source, Gigahertz Peaked Spectrum (GPS) sources
and Compact Symmetric Objects (CSO), are the  
most likely representatives of this early evolutionary phase.
\begin{figure}
  \begin{center}
 \includegraphics[width=16cm]{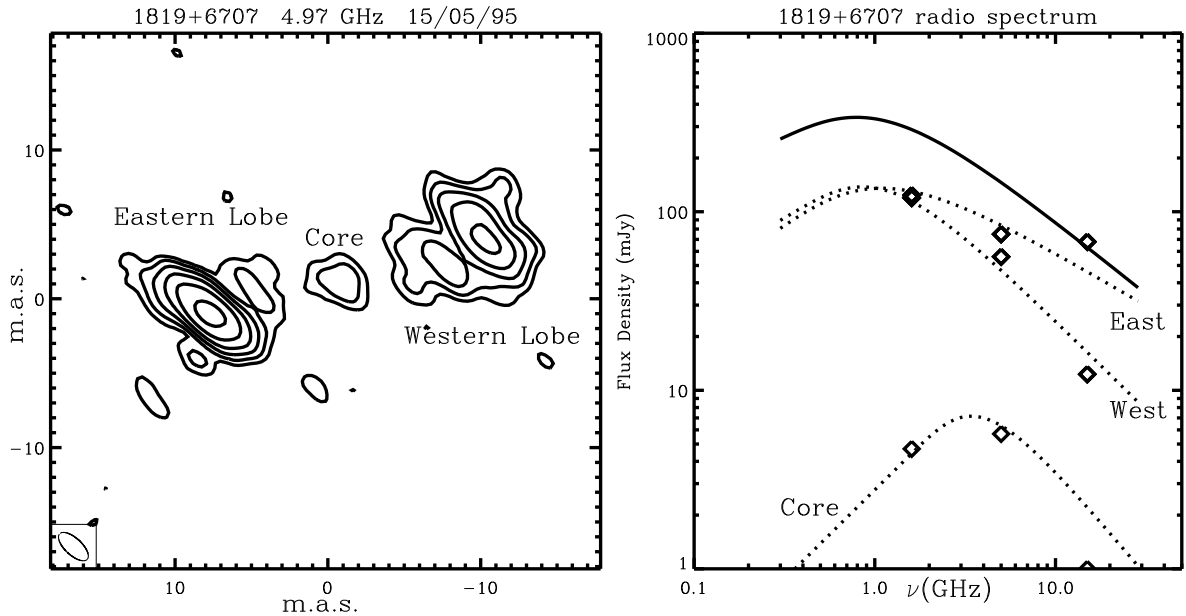}
\end{center}
\label{snelfig1}
\caption{The milli-arcsecond morphology and gigahertz-peaked 
spectrum of the archetype young radio source B1819+6707~\cite{sne97}. 
This radio galaxy (z=0.22) is classified as a CSO since it is compact 
($\sim 100$ pc) and shows structure on both sides of a central core. 
The synchrotron self-absorbed mini-lobes are the main contributors to the 
overall spectrum, causing the turnover at $\sim$1 GHz.}
\end{figure}
GPS sources are characterised by a convex shaped radio spectrum peaking 
at about 1 GHz in frequency~\cite{ode98}, and are selected on the basis of  
interferometric or single dish flux density measurements at several ($>4$) 
frequencies.
CSOs are characterised by their small size ($<500$ pc) and two-sided radio 
structure, e.g.
having jets/lobes on both sides of a central core~\cite{wil94}. 
They are selected on their
milli-arcsecond morphology, which require high resolution VLBI observations 
at, at least, 2 frequencies.
Since CSO and GPS sources are selected in such different ways, studies
of these objects have mostly been presented separately. However, a 
significant overlap between these two classes exists.
GPS sources optically identified with galaxies are most likely to
possess compact symmetric morphologies~\cite{sta98}, and the large majority of 
CSOs exhibit
a gigahertz-peaked spectrum. The large but not complete overlap between
these two classes of sources is caused by the synchrotron self-absorbed 
mini-lobes, located at the extremities of most CSOs, being the main 
contributors to the overall radio spectrum, and producing 
the peak at about 1 GHz in frequency (fig \ref{snelfig1}).
Since orientation effects can influence both 
the observed radio morphology and the radio spectrum, the selection of
an object as a CSO or GPS source depends on its viewing angle~\cite{sne98a}. 
This causes
the overlap to be less than 100\%. Furthermore, 
GPS sources optically identified 
with quasars are preferentially found to have core-jet 
morphologies~\cite{sta98}.
The morphological dichotomy of GPS galaxies and quasars and their very 
different redshift distributions, make it unlikely that GPS galaxies 
and quasars are related by orientation; they may just happen to have similar
radio spectra~\cite{sne99}.

The study of complete samples of young objects can constrain the 
evolution of radio sources, e.g. by using their number counts and source
size distributions as function of flux density. The selection effects in these
samples have to be understood to be able to compare these 
statistics with those for old, large size radio sources.
Radio surveys at generally two frequencies are needed to select 
GPS candidates. In this process strong selection effects are evident in
peak frequency and peak flux density. Furthermore, the boundary 
between a spectrum to be peaked or not-peaked, eg. its curvature, is 
drawn quite arbitrarily. Since the selection surveys and additional 
observations
are most likely to have taken place at different epochs, variability
may also play a role in the selection. 
The selection of a complete sample of CSOs needs at least VLBI surveys
at two frequencies (eg. the Caltech-Jodrell Bank I\&II 
Surveys~\cite{pol95,tay95}).
Dependent on the resolution and dynamic range of the VLBI observations at 
all observing frequencies, strong selection effects are made on the overall
angular size of the source, the contrast of the core to the approaching and 
receding 
sides of the source, and the spectral indices  of the different components
which are required to establish the two-sided nature of the radio morphology.
For example, if observations with infinite dynamic range were possible,
all compact core-jet sources  could be classified as CSOs, since their 
counter-jets would be visible. In addition, VLBI surveys are mainly 
completed on samples of flat spectrum sources. In this way, a significant
fraction of CSOs with steeper spectra may be missed.

It is relatively straightforward to select young radio sources on the
basis of their gigahertz-peaked spectrum, while the selection on 
compact symmetric morphology is non trivial, in particular 
for very compact and/or faint sources. If a complete sample of 
young radio sources is required, selection on a gigahertz-peaked spectrum is 
therefore preferable, especially at fainter flux density levels.
However, it is probably preferable to omit those optically identified with 
quasars, since their relation to young sources is doubtful.
For the detailed analysis of individual objects, it is preferable
to use confirmed CSOs, since the nature of their different components 
and possible orientation effects are better understood.

\section{Evidence for GPS/CSO being young}

Since the initial discovery of GPS sources, it has been speculated
that these are young objects~\cite{shk65,bla70}. 
However, a commonly 
discussed alternative to them being young was that they are small 
due to confinement by a particularly dense and clumpy interstellar
medium that impedes the outward propagation of the 
jets~\cite{vbr84,ode91}. 
This latter hypothesis now looks less likely since recent observations show 
that the surrounding media of peaked spectrum sources are not significantly
different from large scale radio sources, and insufficiently dense to 
confine these objects.
More convincingly, the propagation velocities of the hot spots of several CSOs
have now been measured to be $\sim 0.2h^{-1}c$~\cite{ows98,ows98b,tsc},
giving an 
apparent age of $\sim 10^3$ year and clearly showing that these are
indeed young objects.
Recent determinations of the radiative ages from the high frequency breaks 
GPS sources and the larger Compact Steep Spectrum (CSS) sources are 
found to be consistent with ages ranging from $10^3-10^5$ years~\cite{mur99}.

\section{Current Views on Radio Source Evolution}

Observational constraints on the luminosity evolution of radio sources
mainly come from the source density in the power - linear size ($P-D$)
diagram~\cite{shk63}. It was found that sources with large sizes
($D>1$ Mpc) and high radio luminosities ($P>10^{26}$ W/Hz at 178 MHz) 
are rare, suggesting that the luminosity of sources should decrease quickly
with linear sizes approaching 1 Mpc.
Several evolution scenarios have been proposed for young radio sources, 
in which GPS sources subsequently evolve to Compact Steep Spectrum (CSS) 
sources and large-scale doubles~\cite{hod87,fan95,ode97},
 and CSOs evolve in Medium Symmetric Objects (MSO),
and Large Scale Objects (LSO)~\cite{rea96}. 
In these models, the age ratio of large scale to GPS, and 
LSO to CSO is typically $\sim 10^3$. The much larger fraction (say 10\%) 
of GPS and CSOs in radio surveys therefore implies that young radio
sources have to substantially decrease (a factor $\sim$10)
in radio luminosity when evolving to large size radio sources.
This can be explained by a decrease in radiation efficiency with source 
size~\cite{fan95,rea96}
The transition from CSO to MSO does not have to occur at the 
same moment as the transition from GPS to CSS, and 
depends on the quite arbitrary definitions of the different classes of object. 
eg. A CSO can have such a low spectral peak frequency that it is actually 
defined as a CSS and not as a GPS source.

Several young radio sources (eg. 0108+388;~\cite{bau90}) exhibit low
level, steep spectrum, extended emission on arcsecond scales, which seem to
be relics of much older radio-activity. These objects are generally classified
as being intermittent or re-occurent, and not as young objects.
However, the components related to their gigahertz-peaked spectra and
CSO morphologies are certainly young, and we therefore believe it 
is correct to call them young objects. The presence of faint relic emission
only indicates that the active nucleus has been active before, and may
constrain the typical timescale and frequency of such an event. 
Based on the current knowledge of the formation of massive 
black-holes in the centers of galaxies (eg.~\cite{ric98}), 
it is unlikely that the central engine itself is young, but only the 
radio source. 

It is unclear whether all young sources actually evolve to large extended
objects. Some, or even the majority, may be short-lived phenomena due
to a lack of significant fuel~\cite{rea94}. The possible existence of 
these objects can largely influence the source statistics of young radio
sources, and their luminosity evolution.

\section{GPS sources at faint flux densities}

In addition to the GB6 \cite{con91} survey at 5 GHz, several new surveys 
have become available in recent years, like the WENSS at 325 
MHz~\cite{ren97}, and the NVSS~\cite{con99} and FIRST~\cite{bec99} at 1.4 GHz.
 These surveys form a very powerful combination
to select large and homogeneous samples of GPS candidates at faint 
flux density levels.
The study of GPS samples at faint and bright flux density levels 
allow a disentanglement of redshift and radio luminosity effects.
A small sample of 47 faint GPS sources has been investigated 
by our group, which was selected from first available areas of the WENSS 
survey~\cite{sne98b}
The sample has been studied extensively in the optical to determine
the nature and redshifts of the optical identifications, resulting in an 
identification fraction of $87\%$~\cite{sne98c}. 
About 40\% of the sample consists of 
high redshift quasars (which we will further ignore). Only a few of 
the redshifts of GPS galaxies have been determined yet, due to their faint
magnitudes and weak emission lines~\cite{sne99}. Fortunately their redshifts 
can be estimated due to their well established Hubble diagram~\cite{sne96}.
Global VLBI observations at 5 GHz were obtained for all sources in the sample.
In addition, observations at 1.6 and 15 GHz with the global array and the VLBA 
respectively were taken. In this way, 94\% percent of the sources in the 
sample were observed at least at two frequencies, above and at or below their 
spectral peak~\cite{sne97}.

The combination of this faint GPS sample, and bright GPS and CSS samples
from the literature~\cite{sta98,fan90} gave a 
unique opportunity to investigate the relation between spectral peak and 
size of young radio sources. Not surprisingly, the well-known correlation
between peak frequency and angular size~\cite{fan90} was confirmed. 
However, in addition, a correlation was found between the peak flux density 
and angular size. 
Most remarkably, the strength and signs of these two correlations are
exactly as expected for synchrotron self absorption (SSA). This strongly 
suggests that SSA is indeed the cause of the spectral turnovers in GPS and CSS
sources, and not free-free absorption as recently proposed by~\cite{bic97}.
The spectral peak originates in the dominant features of the 
radio source, the mini-lobes, and 
therefore reflects the sizes of the mini-lobes. The angular size 
from the VLBI observations is the overall size of the radio source, eg. the 
distance between the two mini-lobes. The correlations
between the spectral peak and size therefore imply a linear correlation 
between the mini-lobes and overall sizes, meaning that during the evolution 
of young radio sources the ratio of the size of the mini-lobes and the 
distance between the two mini-lobes is constant. This suggests they evolve in 
a self-similar way.

\section{Luminosity Evolution and the Luminosity Function}

\begin{figure}
  \begin{center}
 \includegraphics[width=8cm]{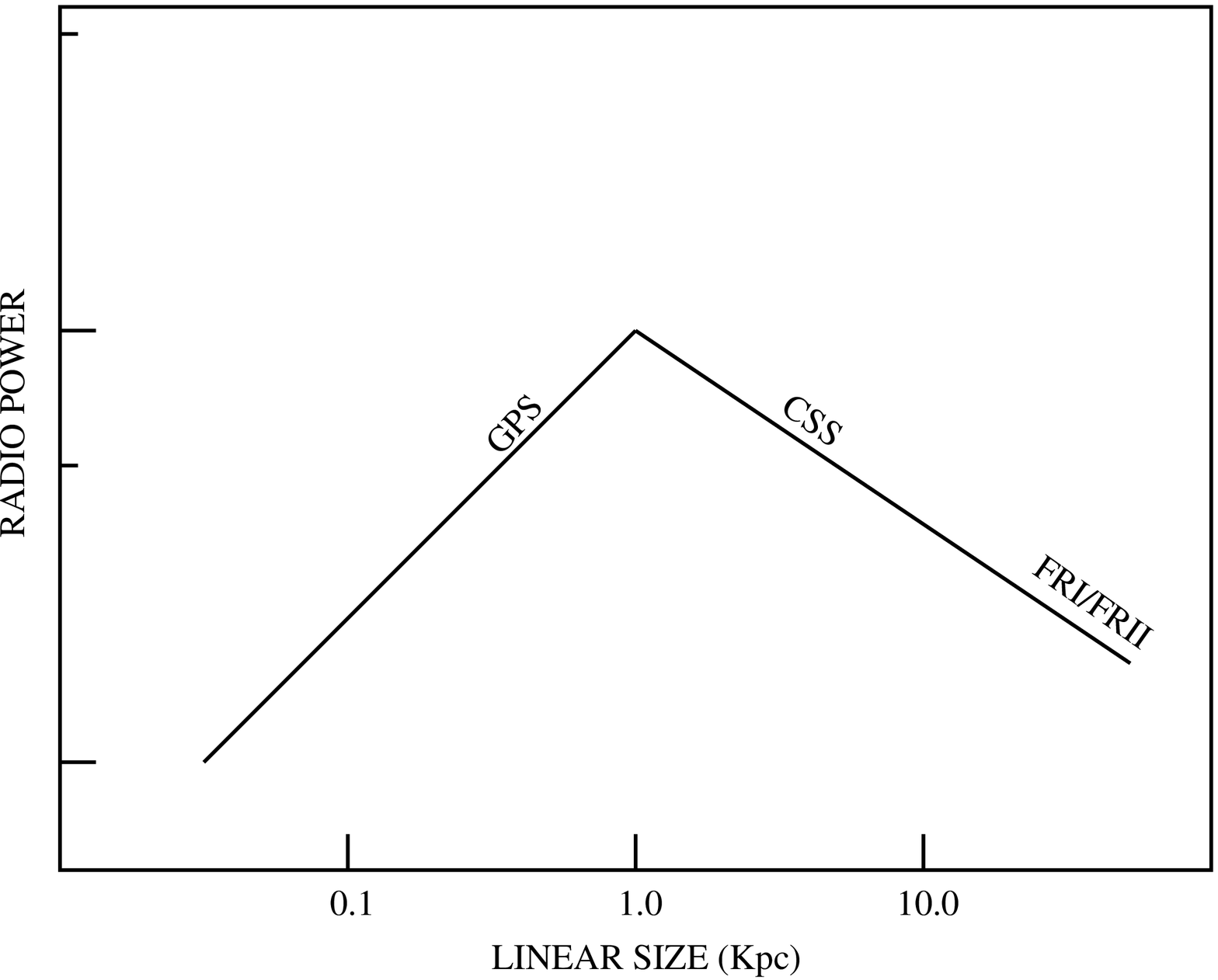}
 \includegraphics[width=7cm]{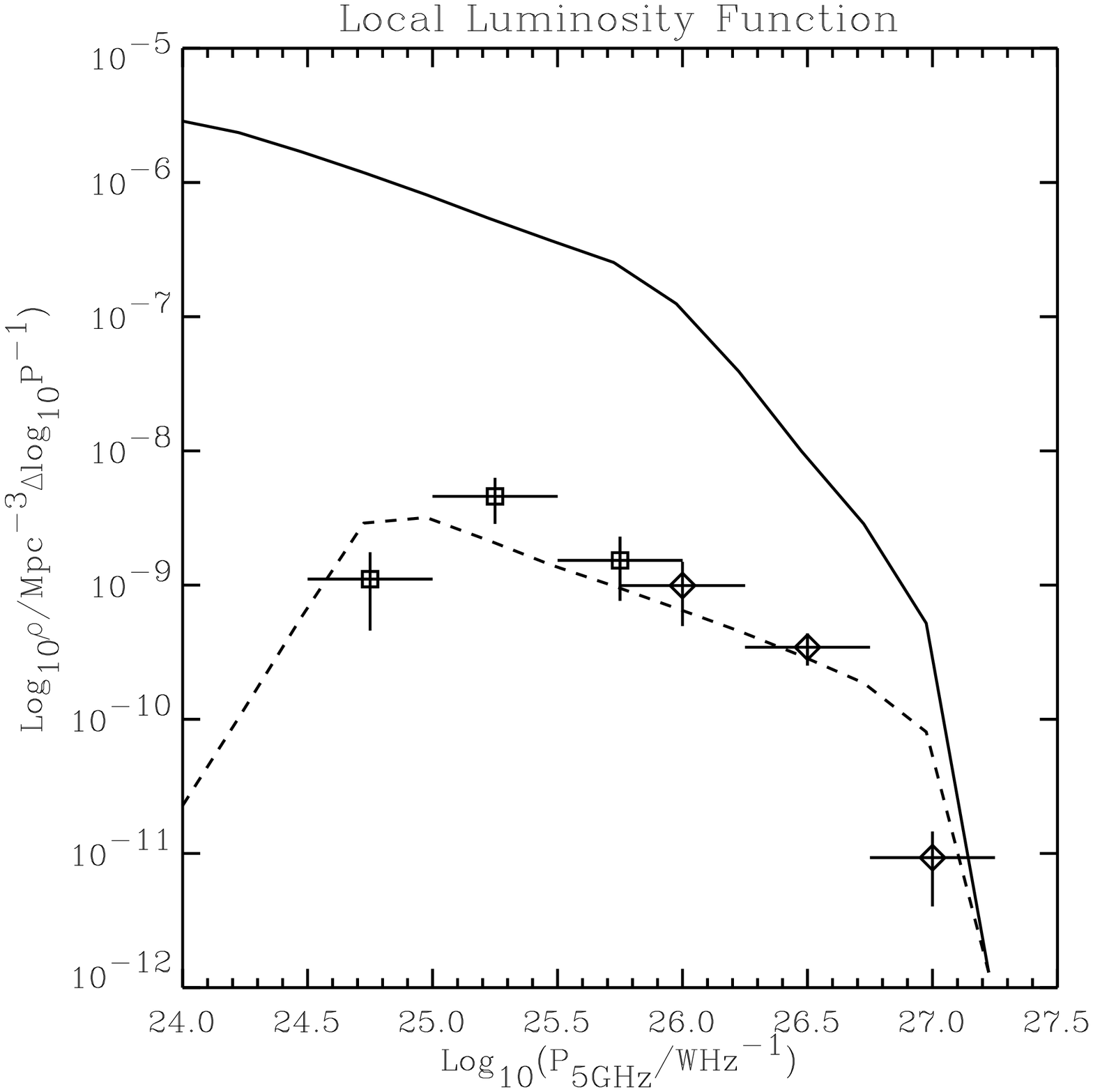}
\end{center}
\caption{(Left) The proposed evolution scenario in which young radio sources
increase in radio luminosity and large extended objects decrease in luminosity
with time. (right) The local luminosity function (LLF) as derived from the 
bright Stanghellini
et al.~\cite{sta97} and faint GPS samples using the cosmological number 
density evolution as determined for steep spectrum radio sources. 
The solid and dotted lines are simulated LLFs for old and young radio sources
respectively.}
\label{llf}
\label{le}
\end{figure}

In flux density limited samples, GPS galaxies are found at higher 
redshifts than large size radio sources~\cite{sne97}. Since 
the lifetimes of radio sources are short compared to cosmological timescales,
this can only mean that the slope of their luminosity functions
are different, if GPS sources are to evolve to large size radio sources.
We argue that the slope of the luminosity function is strongly
dependent on the evolution in radio power of the individual sources. 
To explain the difference in redshift distribution, we propose a 
luminosity evolution scenario in which GPS sources increase in luminosity
and large extended objects decrease in luminosity with time (Fig. \ref{le}).
Sources in a volume-based sample are biased towards low jet-powers and older 
ages, for populations of both GPS and extended objects. 
Low jet powers result in low luminosity sources. The higher the age 
of a large scale source the lower its luminosity, but the higher the age of
a GPS source the higher its luminosity. This means that for a population of 
large scale sources the jet power and
age biases strengthen each other resulting in a steep luminosity function,
while they counteract for GPS sources, resulting in a flatter luminosity 
function. 
The evolution scenario proposed is expected for a ram-pressure confined 
radio source in a surrounding medium with a King profile density.
In the inner parts of the King profile, the density of the medium is constant 
and the radio source builds up its luminosity, but after it grows large enough 
the density of the surrounding medium declines and the luminosity of the radio
source decreases.

Triggered by the ideas above, a new method has been developed to constrain
the luminosity evolution of radio sources by comparison of 
the local luminosity functions (LLF) of young and old objects.
At present, an insufficient number of GPS sources are known at low redshift 
to construct an LLF directly. However, the cosmological 
number density evolution, as derived for steep spectrum (eg. large size) radio 
sources by~\cite{dun90}, is used to derive a LLF for young 
radio sources from the GPS samples. 
The result, as is shown in Fig. \ref{llf} is consistent with the luminosity scenario as proposed above. Note however, that the faint and bright GPS 
samples were selected in different ways resulting in large uncertainties.

\section{The Future and the Square Kilometer Array}

The new large area radio surveys and the current sensitivity and 
flexibility of VLBI networks allow the construction and investigation of 
large and homogeneous samples of young sources. 
This is fueling the rapid development of this research area, with the 
exciting prospect this continuing to do so for the next few years at least.
The total number of GPS sources in the WENSS survey selectable on the basis
of their inverted spectra between 325 MHz and 1.6 GHz (NVSS),
is likely to be on the order of $2\times10^3$, from which 
about $100-200$ can be identified with low redshift galaxies. 
This will allow a direct determination of the local luminosity 
function for young radio sources, down to a 5 GHz radio power of 
$10^{24}$ W/Hz. 

A strong impact on the research area of compact radio sources can be expected 
from the planned Square Kilometer Array (SKA), in particular if the 
configuration includes multi- $10^3$ km baselines, or if it is added to 
ground and space VLBI networks. The unrivaled image quality provided by its
quasi-continuous uv-coverage and its high sensitivity combined with 
m.a.s. resolution, promises to give new insights into the physics of 
(relativistic) jets (eg. Kirchbaum et al., this volume).
The largest contribution of SKA to the statistical properties of young 
radio sources, as discussed in this paper, can be expected from its ability
to select and investigate weak ($P_{5 GHz}\sim10^{24}$ W/Hz) young radio 
sources
out to much larger cosmological distances ($z\sim2$ instead of $z\sim 0.1$),
allowing a detailed comparison of the cosmological number density 
evolution of young to that of old sources over a wide range of luminosity.
This will put much stronger constraints on 
the luminosity evolution of the individual objects and it will provide
new insights into the strong cosmological evolution
of radio sources from high redshift to the present.

\section*{Acknowledgements}
This research was in part funded by the European Commission under
contract  ERBFMRX-CT96-0034 (CERES)

\end{document}